\documentclass[journal=jpcafh,manuscript=article]{achemso}
\usepackage{amsmath}
\usepackage{amssymb}
\usepackage[version=4]{mhchem}
\usepackage{xcolor}
\usepackage{hyperref}
\usepackage{mathtools,xparse}
\usepackage{placeins}
\usepackage{xr}
\makeatletter
\DeclareFontFamily{OMX}{MnSymbolE}{}
\DeclareSymbolFont{MnLargeSymbols}{OMX}{MnSymbolE}{m}{n}
\SetSymbolFont{MnLargeSymbols}{bold}{OMX}{MnSymbolE}{b}{n}
\DeclareFontShape{OMX}{MnSymbolE}{m}{n}{
    <-6>  MnSymbolE5
   <6-7>  MnSymbolE6
   <7-8>  MnSymbolE7
   <8-9>  MnSymbolE8
   <9-10> MnSymbolE9
  <10-12> MnSymbolE10
  <12->   MnSymbolE12
}{}
\DeclareFontShape{OMX}{MnSymbolE}{b}{n}{
    <-6>  MnSymbolE-Bold5
   <6-7>  MnSymbolE-Bold6
   <7-8>  MnSymbolE-Bold7
   <8-9>  MnSymbolE-Bold8
   <9-10> MnSymbolE-Bold9
  <10-12> MnSymbolE-Bold10
  <12->   MnSymbolE-Bold12
}{}

\let\llangle\@undefined
\let\rrangle\@undefined
\DeclareMathDelimiter{\llangle}{\mathopen}%
                     {MnLargeSymbols}{'164}{MnLargeSymbols}{'164}
\DeclareMathDelimiter{\rrangle}{\mathclose}%
                     {MnLargeSymbols}{'171}{MnLargeSymbols}{'171}
\makeatother

\newcommand{\ket}[1]{ \left| #1 \right \rangle}

\author{Peter Reinholdt}
\affiliation[SDU]{Department of Physics, Chemistry and Pharmacy, University of Southern Denmark, Campusvej~55, DK--5230 Odense M, Denmark}
\email{reinholdt@sdu.dk}

\author{Erik Kjellgren}
\affiliation[SDU]{Department of Physics, Chemistry and Pharmacy, University of Southern Denmark, Campusvej~55, DK--5230 Odense M, Denmark}

\author{Jacob Kongsted}
\affiliation[SDU]{Department of Physics, Chemistry and Pharmacy, University of Southern Denmark, Campusvej~55, DK--5230 Odense M, Denmark}

\title{Linear Response Selected Configuration Interaction}

\begin{document}

\begin{tocentry}
\includegraphics[width=8.25cm]{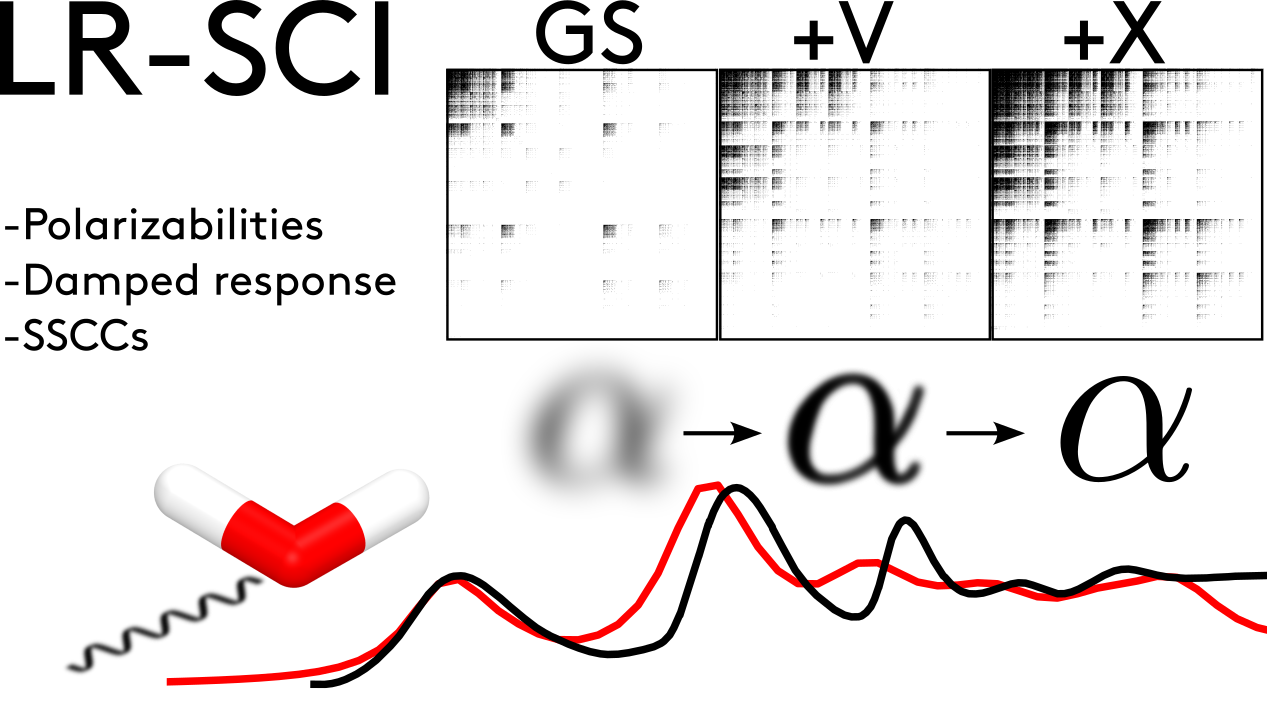}
\end{tocentry}

\begin{abstract}
In this work, we extend selected configuration interaction (SCI) methods beyond energies and expectation values by introducing a linear response (LR) framework for molecular response properties. Existing SCI approaches are capable of approximating the energy of the full configuration interaction (FCI) wave function with high accuracy, but at a much lower cost.
However, conventional determinant selection will, by design, mainly select determinants that are expected to improve energies, and this can lead to the omission of many determinants that are important for wave function response.
We address this by introducing two new selection criteria motivated by linear response theory.
Using these extended determinant selection criteria, we demonstrate that LR-SCI can systematically converge toward the FCI limit for static polarizabilities. 
Using a damped LR formulation, we compute the water K-edge X-ray absorption spectrum in active spaces up to (10e, 58o). 
Finally, we use LR-SCI to compute NMR spin-spin coupling constants for water, where we find that accuracy beyond that offered by CCSDT can be achieved.
Overall, LR-SCI offers a promising route to compute response properties with near-FCI accuracy to systems beyond the reach of exact FCI.
\end{abstract}

\section{Introduction}
The full configuration interaction (FCI) method provides the exact solution to the electronic Schr\"odinger equation within a given basis. Due to an inherent exponential computational scaling with respect to the system size, exact FCI is limited to only quite small molecular systems.
Even so, through a combination of hardware and algorithmic advances, FCI has been pushed to relatively large-scale CI calculations with trillions\cite{gao2024distributed} or even quadrillions of determinants\cite{shayit2025breaking}, such as the widely noted work by \citeauthor{gao2024distributed}\cite{gao2024distributed} on propane/STO-3G (26e, 23o).
However, quasi-exact energies can be obtained much more affordably using approximate (yet highly accurate) FCI methods, as was pointed out by \citeauthor{loos2024go}\cite{loos2024go} and later \citeauthor{craciunescu2025selected}\cite{craciunescu2025selected}

Several computational methods with significantly different theoretical foundations have demonstrated near-exact FCI results for system sizes well beyond the reach of exact FCI, including ones based on the density matrix renormalization group (DMRG)\cite{chan2011density}, many-body expanded FCI (MBE-FCI)\cite{eriksen2017virtual,eriksen2018many,eriksen2019many}, and selected configuration interaction (SCI)\cite{gershgorn1968application,huron1973iterative,caballol1992direct,tubman2016deterministic,holmes2016heat,liu2016ici}, among others\cite{motta2018ab,xu2018full,li2022downfolded}.
For the most part, such quasi-exact methods are focused on obtaining accurate energies, especially the ground-state energy.

In this work, we focus on SCI-type methods, which are configuration interaction (CI) approaches that iteratively construct the wave function by identifying and including only the most important determinants, exploiting the sparsity of the CI vector to yield compact and systematically improvable approximations to FCI.
Excited states can be obtained with SCI by explicitly solving for several (low-lying) eigenstates, which allows for the evaluation of highly accurate excitation energies as energy differences between different approximate eigenstates\cite{loos2018mountaineering}.
There is thus no need for a linear response framework for obtaining excitation energies with SCI. 
Beyond energies, expectation values, and transition moments between different eigenstates are also straightforward to obtain, giving access to, e.g., dipole moments, oscillator strengths, and hyperfine coupling constants\cite{angeli1998multireference,angeli2001multireference,damour2022ground}.
However, treating molecular properties beyond expectation values or transition moments has, so far, not been explored in detail for SCI-type methods.

Some static properties, such as polarizabilities, can (in principle) be obtained from finite-field calculations. However, a proper LR formulation provides access to frequency-dependent and damped response functions, enabling properties such as core-excited spectra\cite{ekstrom2006x,fahleson2016polarization,fransson2016k} and C$_6$ dispersion coefficients\cite{fowler1990c,jiemchooroj2005complex}.
Many important molecular properties are naturally formulated within response theory \cite{norman2011perspective,helgaker2012recent}, including polarizabilities, magnetizabilities, NMR shielding tensors, and spin-spin coupling constants. A linear response framework thus greatly extends the scope of accessible molecular properties and spectroscopic observables.
Although we are not aware of any existing LR-SCI implementations, we note that there has been some work reported on (real) time-propagation with the time-dependent adaptive configuration interaction (TD-ACI) method\cite{schriber2019time} and time-dependent adaptive sampling configuration interaction (TD-ASCI) method\cite{shee2025real}, which gives access to absorption spectra by Fourier transforming the time-domain correlation functions.
Linear response theory has also been explored within other theoretical frameworks capable of providing near-FCI results, notably with DMRG\cite{dorando2009analytic,nakatani2014linear}, and for FCIQMC\cite{booth2012communication,blunt2015krylov,samanta2018response}.

\begin{figure}
    \centering
    \includegraphics[width=0.7\linewidth]{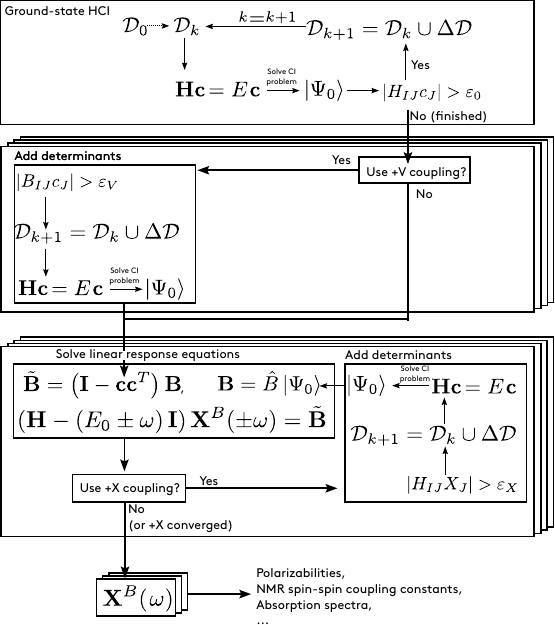}
    \caption{Schematic overview of the LR-SCI calculation. The upper box shows the ground-state HCI iterations, where the ground-state CI problem is solved and the determinant space $\mathcal{D}_k$ is iteratively expanded by adding determinants $\Delta \mathcal{D}$ that satisfy the criterion in Eq. \eqref{eq:hci_criterion} until convergence. The middle box shows the additional property-gradient determinant addition step, which is included with the +V coupling (Eq. \eqref{eq:hci_prop_criterion}). The lower box shows the formation of the property gradient and the solution of the linear response equations. If the +X coupling (Eq. \eqref{eq:hci_rsp_criterion}) is used, an iterative determinant-addition step is included to identify determinants that couple strongly to the response vector. Finally, the converged response vectors are used to evaluate molecular properties, such as polarizabilities or NMR spin-spin coupling constants.}
    \label{fig:lrsci_workflow}
\end{figure}

In this work, we extend SCI methods to molecular response properties using a variant of Heat-bath CI (HCI)\cite{holmes2016heat}.
Applying linear response theory with SCI wave functions is formally straightforward, as the method is CI-based and standard exact-state theory applies. 
One of the main challenges lies in determinant selection, as determinants important for response properties are often different from those that are required for an accurate description of the ground state.
Ultimately, this can lead to slow or even non-convergent response functions as the CI expansion grows.
We note that related considerations for ground-state expectation values have been addressed with the \emph{property}-focused selection scheme by \citeauthor{angeli2001multireference}\cite{angeli2001multireference}.
To address this, we introduce response-theory motivated determinant-selection criteria to augment an initial ground-state SCI wave function with determinants expected to be relevant for molecular response properties, and demonstrate the approach for a range of properties, including static polarizabilities, damped polarizabilities applied to X-ray absorption spectra, and nuclear spin-spin coupling constants.
The overall workflow of the LR-SCI procedure, including the iterative selection of determinants for both the ground and response spaces, is illustrated in Figure \ref{fig:lrsci_workflow}.

\section{Theory}
\subsection{Selected Configuration Interaction}
A configuration interaction (CI) wave function can be expressed as a linear combination of Slater determinants $\ket{\Phi_I}$
\begin{equation}
    \ket{\Psi} = \sum_{I} c_I \ket{\Phi_I},
\end{equation}
with the CI coefficients, $c_I$. 
The coefficients are determined by solving the time-independent electronic Schr\"odinger equation,
\begin{equation}
\hat{H} \ket{\Psi_k} = E_k \ket{\Psi_k},
\end{equation}
which, when projected onto the determinant basis, leads to the matrix eigenvalue problem
\begin{equation}
\mathbf{H} \mathbf{c}_k = E_k \mathbf{c}_k.
\end{equation}
The energy $E_k$ is an eigenvalue of the Hamiltonian matrix with elements $H_{IJ} = \langle \Phi_I | \hat{H} | \Phi_J \rangle$, and $c_I$ are the coefficients of the corresponding eigenvector.

The included determinants can be selected through various means. 
If all possible determinants are included, one arrives at the full configuration interaction (FCI) wave function, which gives the exact solution to the electronic Schr\"odinger equation within a given basis. However, due to the combinatorial scaling of placing $N$ electrons in $M$ orbitals, the number of determinants increases rapidly with increasing molecular size, limiting exact FCI to only very small systems. 
The CI expansion can be truncated by excitation level out of a reference determinant, forming the CISD, CISDT, etc. hierarchy\cite{siegbahn1992configuration}, which leads to a polynomially scaling hierarchy. However, truncated CI methods lack size-consistency\cite{bartlett1977determination} and can include many determinants with vanishing CI coefficients that are not relevant for the CI wave function\cite{ivanic2001identification}.
In SCI expansions \cite{huron1973iterative,tubman2016deterministic,holmes2016heat,liu2016ici}, this sparsity is exploited, and various heuristics are used to select a relatively small set of important determinants, which ideally should give compact yet accurate expansions.

In this work, we will consider Heat-bath Configuration Interaction\cite{holmes2016heat} (HCI), where determinants for the CI expansion are included based on the criterion
\begin{equation}
    \left|H_{IJ} c_J \right| > \varepsilon_0. \label{eq:hci_criterion}
    \end{equation}
Here, $H_{IJ}$ is the Hamiltonian matrix element between an already included determinant $\ket{\Phi_J}$ and the candidate determinant $\ket{\Phi_I}$, while $c_J$ is the coefficient of the included determinant $\ket{\Phi_J}$. This selection focuses on including determinants with significant coupling to the existing CI expansion, typically yielding compact and accurate CI expansions.

The HCI method works iteratively, starting from some initial CI expansion:
1) the CI Hamiltonian is diagonalized, and the CI coefficients of the current determinant set are obtained;
2) determinants are added according to the criterion in Eq. \eqref{eq:hci_criterion}, with 
$\varepsilon_0$ as a parameter.
These steps are repeated until no more determinants can be added according to Eq. \eqref{eq:hci_criterion} (in the original work\cite{holmes2016heat}, when the number of new determinants is less than 1\% of the number of determinants already selected). 

\subsection{Linear Response Theory}
Following \citeauthor{koch1991analytical}\cite{koch1991analytical}, frequency-dependent linear response properties for a CI wave function can be obtained by solving two sets of linear response equations 

\begin{equation}
    \left( \mathbf{H} - \left(E_0\pm\omega\right)\mathbf{I}\right) \mathbf{X}^B(\pm \omega) = \tilde{\mathbf{B}}, \label{eq:linear_response_equation}
\end{equation}
where the response vector $\mathbf{X}^B(\pm \omega)$ is the solution to the linear response equation, $\omega$ is the perturbation frequency, and $E_0$ is the ground-state energy of $\ket{\Psi_0}$. 
The property gradient $\tilde{\mathbf{B}}$ of the one-electron operator $\hat{B}$ is defined 
\begin{equation}
    \tilde{\mathbf{B}} = \left(\mathbf{I} -\mathbf{c}\mathbf{c}^T\right)\mathbf{B}, \label{eq:property_gradient}
\end{equation}
where
\begin{equation}
    B_j = \left<\Phi_j \left|\hat{B}\right|\Psi\right>.
\end{equation}
The property gradient corresponds to the action of the one-electron operator $\hat{B}$ on the CI wave function, projected to remove the ground-state contribution.
Linear response functions (e.g., dipole--dipole polarizabilities) can be evaluated using the response- and property vectors as 
\begin{equation}
    -\left\llangle\mu^A;\mu^B\right\rrangle_\omega = \alpha_{AB}(\omega) = \tilde{\mathbf{A}}^T \left(\mathbf{X}^B(\omega) + \mathbf{X}^B(-\omega) \right),
\end{equation}
where $\tilde{\mathbf{A}}$ is the property gradient of the one-electron operator $\hat{A}$ (see Eq. \eqref{eq:property_gradient}). 
The solution of the linear response equations can be done efficiently using iterative schemes such as Davidson-type methods\cite{DAVIDSON197587,olsen1988solution}, in which one only requires the ability to form matrix-vector products of the CI Hamiltonian with some arbitrary trial vector.
Thus, the explicit construction (and inversion) of the full $N_\mathrm{det}\times N_\mathrm{det}$ Hamiltonian matrix is not required.

The response functions from regular frequency-dependent linear response theory have divergences (poles) at frequencies corresponding to excitation energies (i.e., when $\mathbf{H} - \left(E_0\pm\omega\right)\mathbf{I}$ is singular).
However, the theory can be extended to a framework that is convergent at all frequencies\cite{norman:2001cpplinear,norman:2005nonlinear,kristensen2009quasidampedresponse,norman2011perspective,helgaker2012recent,kauczor2013dampedcc,faber2019:rixs} by replacing the real frequency $\omega$ with the complex variable $\omega + i\gamma$.
The parameter $\gamma$ introduces a finite lifetime for the excited states, effectively broadening the discrete excitation poles into Lorentzian line shapes. In this way, $\gamma$ serves as a phenomenological linewidth that mimics relaxation and decay processes.
This ensures that the response functions remain finite across all frequencies, with the response function becoming purely imaginary at frequencies corresponding to excitation energies.
The Davidson method can be carried out without issue in complex algebra, which is the approach we have used for our implementation.
One of the important properties which can be accessed with a damped linear response theory framework is the absorption cross-sections, which can be extracted from the imaginary part of the complex dipole--dipole polarizability as\cite{fransson2016k}
\begin{equation}
    \sigma(\omega) = \frac{\omega}{\epsilon_0 c} \mathrm{Im}(\alpha_{\mathrm{iso}}(\omega)),
    \label{eq:sigma_abs}
\end{equation}
where the isotropic polarizability is the average of the diagonal components, $\alpha_\mathrm{iso} = \frac{1}{3} (\alpha_{xx}  + \alpha_{yy} + \alpha_{zz})$.

\subsection{Spin-spin coupling constants}
The experimentally measurable indirect nuclear spin-spin coupling constant ($J$) between two nuclei $A$ and $B$ can be related to the reduced indirect nuclear spin-spin coupling constant ($K$) as
\begin{equation}
    J_{AB} = \frac{\mu_N^2 g_A g_B}{h}  K_{AB},
\end{equation}
where  $g_A$ and $g_B$ are the (isotope-dependent) nuclear $g$-factors, $h$ is the Planck constant, and $\mu_N$ is the nuclear magneton. 
The reduced indirect nuclear spin-spin coupling can be calculated as\cite{spasB8,helgaker2008quantum} 
\begin{equation}
    K_{AB,\alpha\beta} = \left<\hat{O}^{\mathrm{DSO}}_{AB,\alpha\beta}\right>
    + \left\llangle\hat{O}^{\mathrm{PSO}}_{A,\alpha};\hat{O}^{\mathrm{PSO}}_{B,\beta}\right\rrangle
    + \left\llangle\hat{O}^{\mathrm{FC}}_{A,\alpha};\hat{O}^{\mathrm{FC}}_{B,\beta}\right\rrangle
    + 2\left\llangle\hat{O}^{\mathrm{FC}}_{A,\alpha};\hat{O}^{\mathrm{SD}}_{B,\beta}\right\rrangle
    + \left\llangle\hat{O}^{\mathrm{SD}}_{A,\alpha};\hat{O}^{\mathrm{SD}}_{B,\beta}\right\rrangle
\end{equation}
where $\alpha$ and $\beta$ refer to Cartesian components. 
The reduced indirect nuclear spin-spin coupling constant can thus be evaluated from an expectation value of the diamagnetic spin-orbit (DSO) operator,
\begin{equation}
    \hat{\boldsymbol{h}}^{\mathrm{DSO}}_{AB} = \alpha^4 \sum_i \frac{\boldsymbol{r}_{iA}^T \boldsymbol{r}_{iB} \boldsymbol{I} - \boldsymbol{r}_{iA} \boldsymbol{r}_{iB}^T}{r_{iA}^3 r_{iB}^3}
\end{equation}
and from a set of response functions, namely those of the  paramagnetic spin-orbit operator (PSO, imaginary singlet perturbation), defined as
\begin{equation}
    \hat{\boldsymbol{h}}^{\mathrm{PSO}}_{A} = \alpha^2 \sum_i \frac{\boldsymbol{r}_{iA} \times \boldsymbol{p}_i}{r_{iA}^3},
\end{equation}
of the Fermi contact operator (FC, real triplet perturbation), defined as
\begin{equation}
    \hat{\boldsymbol{h}}^{\mathrm{FC}}_{A} = -\frac{g_e4\pi}{3}  \alpha^2\sum_i \delta(\boldsymbol{r}_{iA}) \boldsymbol{s}_i
\end{equation}
and of the spin-dipolar operator (SD, real triplet perturbation), defined as
\begin{equation}
    \hat{\boldsymbol{h}}^{\mathrm{SD}}_{A} = -\frac{1}{2}g_e\alpha^2 \sum_i \frac{3\boldsymbol{r}_{iA} \boldsymbol{r}_{iA}^T - r_{iA}^2 \boldsymbol{I}}{r_{iA}^5} \boldsymbol{s}_i,
\end{equation}
where $g_e$ is the electron $g$-factor, and $\alpha$ is the fine-structure constant.
There are two commonly adopted conventions for the value of $g_e$ for the FC and SD integrals. Some implementations (e.g., CFOUR) use a value of $-2$, as predicted by the Dirac equation, while other implementations (e.g., Dalton) use a value of $-2.00231930436182$, consistent with the experimental electron $g$-factor. We have adopted the latter convention and have translated the results obtained with CFOUR to this convention.

\subsection{Response-Oriented Determinant Selection}
In principle, the determinant space from a ground-state SCI calculation could be used directly to solve the response equations, but as we shall later show, this approach does not offer robust convergence.
The set of determinants that are important for a good description of the ground-state 
will likely omit a significant fraction of the determinants that are important for a particular response property.
This is not a flaw, but rather a signature that the ground-state selection procedure is working as intended.

When forming the property gradient, the one-electron operator will create singly excited determinants out of the ground state. These contributions may accumulate in determinants that are not included in the ground state determinant set. 
In some cases, one can even anticipate that \emph{all} such determinants will be missing (i.e., if the ground-state and perturbation operator belong to different point-group symmetries). 
To address this shortcoming, one can include an additional determinant selection step to improve the description of the property gradient.
After the ground-state HCI wave function is obtained, additional determinants are added in an HCI-like selection step according to the one-electron operator ($\hat{B}$) of interest 
\begin{equation}
\left|B_{IJ} c_J \right| > \varepsilon_V, \label{eq:hci_prop_criterion}
\end{equation}
where $B_{IJ}=\left<\Phi_I\left|\hat{B}\right|\Phi_J \right>$ is a matrix element of the one-electron operator in the determinant basis. We will denote the use of this selection scheme as "+V" in the following. 
We note that related criteria have been used within TD-ASCI\cite{shee2025real} and for improving the convergence of ground-state expectation values\cite{angeli2001multireference}.
The determinant addition step should be followed by re-optimization of the CI wave function in the expanded space, since the added determinants may lower the energy, and since wave function gradient terms can otherwise enter the linear response equations.

Solving the response equation may also induce coupling to previously neglected determinants. 
Considering the linear response equation (Eq. \eqref{eq:linear_response_equation}), this can be related to the matrix-vector multiplication of $\mathbf{H}$ with the response vector $\mathbf{X}$. 
Thus, one can consider a determinant addition step where, after the response equations are solved, determinants are added according to the criterion
\begin{equation}
\left|H_{IJ} X_J \right| > \varepsilon_X,\label{eq:hci_rsp_criterion}
\end{equation}
i.e., the standard HCI criterion, but using the coefficients from the response vector in place of the CI coefficients.
We will denote the use of this selection criterion as "+X" in the following.
In practice, we apply this criterion iteratively: the response equations are repeatedly solved, and determinants are added using the coefficients from the response vector. The iteration is terminated when an insignificant number of determinants (say, 1\%) are added in the expansion step.
Again, each addition step should be followed by re-optimization of the CI wave function in the expanded space.

These two additional determinant-selection schemes allow us to define four models for SCI linear response, which we shall term GS (only ground-state HCI selection of determinants), GS+V (additional selection for the property gradient), GS+X (additional selection for the response vector), and GS+V+X (additional selection for both the property gradient and the response vector).
We also refer to Figure \ref{fig:lrsci_workflow} for a graphical overview of the LR-SCI workflow.

\section{Computational Details}\label{sec:computational_details}
Molecular geometries of water and ammonia were optimized using frozen-core CCSD(T)/aug-cc-pCVQZ with the Orca program\cite{orca} (version 6.0.0).
Reference FCI or CASCI calculations were conducted using the Dalton \cite{daltonpaper} and PySCF\cite{pyscf} packages.
CCSD and CCSDT calculations of NMR spin-spin coupling constants were carried out with CFOUR\cite{matthews2020coupled}, with the CCSDT calculations relying on an interface to the MRCC\cite{mester2025overview} program.
Hartree-Fock calculations, and one- and two-electron molecular orbital integrals were obtained with PySCF\cite{pyscf}.
The PyCI library\cite{richer2024pyci} was used for the HCI calculations, providing routines for determinant selection, sparse Hamiltonian construction, and Hamiltonian matrix-vector multiplication. 
Our LR-SCI implementation is available at \url{https://github.com/peter-reinholdt/sci-resp}.
We have used a common value ($\varepsilon$) of the three screening thresholds $\varepsilon_0=\varepsilon_V=\varepsilon_X$ (ground-state, +V, and +X) for all calculations, except for a few test calculations in the Supporting Information. Unless explicitly indicated in the results, we have carried out calculations with a series of increasingly tight $\varepsilon$ (logarithmically spaced 0.25 log-units apart).

\section{Results and Discussion}
\subsection{Static polarizabilities of water and ammonia}
We compute the non-zero components ($\alpha_{xx}$, $\alpha_{yy}$, $\alpha_{zz}$) of the static polarizability of water/cc-pVDZ in a (8e, 23o) active space (frozen-core FCI).
With this active space, FCI spans 78.4 million determinants.
We test the various determinant selection schemes and compare their convergence as a function of the threshold $\varepsilon$. We have used a common value of  $\varepsilon$ for every determinant selection step. In principle, one could apply individual thresholds for the different expansion steps, but based on preliminary test calculations, we have not found any clear advantage in doing so.
The threshold $\varepsilon$ is varied to generate a sequence of increasingly converged SCI calculations.

\begin{figure}
    \centering
    \includegraphics[width=1.0\linewidth]{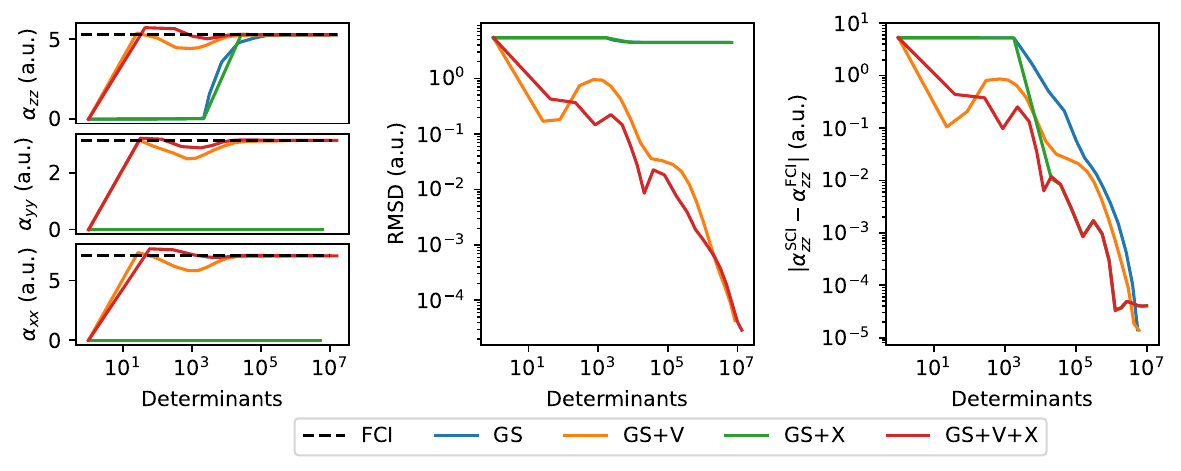}
    \caption{The left panels show the $\alpha_{xx}$, $\alpha_{yy}$, and $\alpha_{zz}$ components of the static polarizability of the water molecule in a cc-pVDZ basis (8e, 23o). The middle panel shows the RMSD between the SCI polarizability and the reference FCI result. The right panel shows the error in $\alpha_{zz}$ relative to the frozen-core FCI reference.}
    \label{fig:water_static_polarizability}
\end{figure}

From Figure \ref{fig:water_static_polarizability}, we see that for the $\alpha_{zz}$ component of the polarizability, all the tested expansion schemes eventually converge towards the reference FCI polarizability (about 5.287 a.u.).
However, the $\alpha_{yy}$ and $\alpha_{xx}$ components can only be correctly described if a coupling to the one-electron operator is included (GS+V or GS+V+X).
This can be understood based on symmetry arguments: the water molecule has $C_{2v}$ symmetry, and the ground-state belongs to the $A_1$ irrep. The Hamiltonian is totally symmetric, and thus does not couple determinants belonging to different irreps, which means that non-$A_1$ determinants will never be added to the SCI wave function using the ground-state selection criterion. 
Determinants in the $B_1$ and $B_2$ irreps are required for describing the $\alpha_{xx}$ and $\alpha_{yy}$ components, which will only be included when a coupling to the one-electron operator is considered. 
The $z$-component of the dipole operator is in the $A_1$ irrep, which means that determinants that are important for the energy may incidentally also be important for describing the polarizability. 

Considering the RMSD across all three components of the polarizability (middle panel), we thus only find convergence towards the FCI reference with GS+V and GS+V+X (orange and red lines).  
The convergence (in terms of the maximum number of determinants required) is broadly similar between the GS+V and GS+V+X schemes, with some advantage to GS+V+X from across the entire region from $2\times 10^2$ to $2\times 10^6$ determinants, after which the two models perform about equally. 
Both methods give a systematic convergence towards the FCI limit. An accuracy of 0.1 a.u. can be achieved with 6,000 (GS+V+X) or 12,000 (GS+V) determinants, while a higher accuracy requirement of 0.01 a.u. requires 136,000 (GS+V+X) or 470,000 (GS+V) determinants.

For the $\alpha_{zz}$-component, all four determinant selection schemes manage to converge towards the FCI limit, and we thus plot this error separately in the right panel of Figure \ref{fig:water_static_polarizability}.
Using only the default Hamiltonian-based HCI selection (GS, blue lines), we observe slower convergence than with some of the more elaborate schemes we have applied. 
The convergence is improved by including a coupling to the property vector (GS+V, orange lines), but convergence to high accuracy remains slow.
By adding a coupling to the response vector (GS+X, green lines, and GS+V+X, red lines), the convergence towards the reference FCI result is significantly improved, and the polarizability can be reproduced with high accuracy. 
For the response-coupled results, we find that as the thresholds are tightened, the inclusion of the property vector coupling becomes inconsequential (green and red lines overlap for a large number of determinants). However, there is clearly some benefit to including a coupling to the property vector in the small-determinant calculation, where the inclusion ensures a qualitatively correct description.

It is interesting to note that there is some oscillatory convergence behavior around the target FCI value (clearly visible for GS+V and GS+V+X in the left panel of Figure \ref{fig:water_static_polarizability}), with convergence starting from below, then overshooting the polarizability, then undershooting, and so on. 
Unlike the energy (which converges from above for variational methods), there is no variational bound on the polarizability.
In the middle and left panels, the oscillatory convergence behaviour is visible through the regular dips towards zero error, which occur when the sign of the error changes.
The lack of monotonic convergence, in contrast to the variational behavior of energies, can be anticipated to make simple extrapolation schemes less straightforward for polarizabilities.
This is a slightly troubling observation, considering that large-scale SCI calculations for energies rarely rely on a purely variational description alone. In practice, large-scale SCI employs a combination of perturbative corrections and extrapolation to achieve near-FCI accuracy.

Response property calculations are often assumed to require a ground-state energy converged to within microhartree accuracy (which corresponds to roughly $10^{-3}$ accuracy in the wave-function gradient) to yield reliable results.
Such a tight convergence criterion becomes difficult to reach for larger systems using purely variational SCI calculations.
However, as shown in the Supporting Information (Figure \ref{fig:si_energy_vs_polarizability}), a relatively loose energy accuracy of $10^{-3}$ Hartree already yields polarizabilities accurate to better than 0.01 a.u., so fortunately, this criterion turns out to be unnecessarily strict in the context of LR-SCI. 
To understand this, it is important to distinguish between determinants contained in the internal determinant space and the complementary external set of determinants. 
The CI gradient within the internal space vanishes when the ground-state CI problem is solved. 
Although the residual CI gradient associated with determinants outside of the internal space is non-vanishing, the external space does not explicitly enter the linear-response equations.
An implicit dependence remains, since the inclusion of the missing determinants would correct the ground-state CI vector, leading to corrections in the property vector and, subsequently, the response vector. 
The external space may also contain many small elements, which, although they individually contribute very little, taken together can lead to appreciable corrections.
Incorporating such contributions through a perturbative treatment could likely enable LR-SCI to be applied to larger systems.

We have generally applied equal thresholds for the ground-state ($\varepsilon_0$) and response-related ($\varepsilon_V$,$\varepsilon_X$) selection thresholds.
It is natural to consider whether it is beneficial to converge the ground-state HCI tightly before attempting any linear response calculations.
As shown in the Supporting Information (see Figure \ref{fig:si_fixed_epsilon} and \ref{fig:si_adjusted_epsilon}), the default choice of equal $\varepsilon$'s results in a decently balanced description (though there is perhaps some room for fine-tuning). Choosing $\varepsilon_0$ significantly tighter than $\varepsilon_V$ and $\varepsilon_X$ degrades performance since in that limit, the +V and +X schemes effectively do nothing, and the regular GS scheme is recovered.

\begin{figure}
    \centering
    \includegraphics[width=1.0\linewidth]{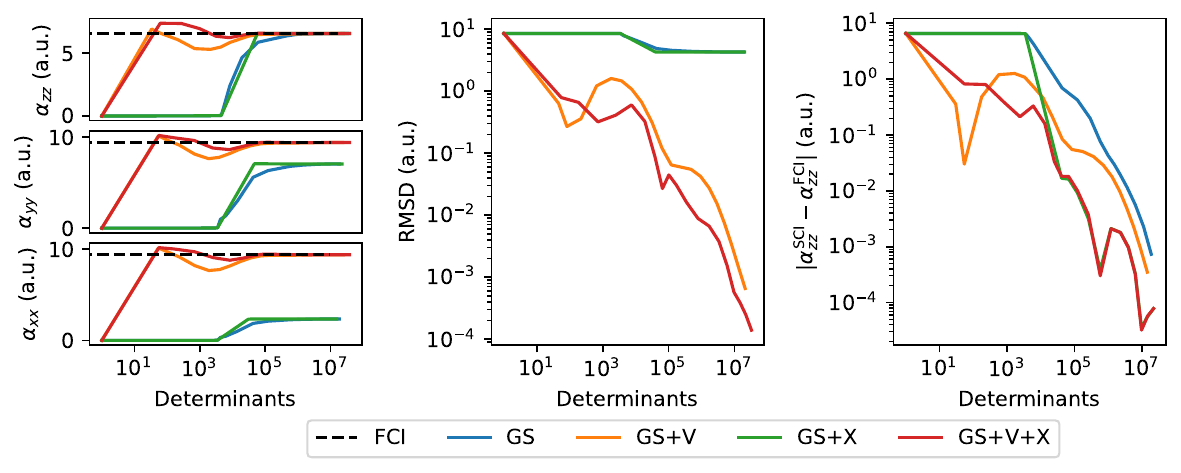}
    \caption{The left panels show the $\alpha_{xx}$, $\alpha_{yy}$, and $\alpha_{zz}$ components of the static polarizability of the ammonia molecule in a cc-pVDZ basis (8e, 28o). The middle panel shows the RMSD between the SCI polarizability and the reference FCI result. The right panel shows the error in $\alpha_{zz}$ relative to the frozen-core FCI reference.}
    \label{fig:ammonia_static_polarizability}
\end{figure}

Next, we consider the polarizability of ammonia/cc-pVDZ, as shown in Figure \ref{fig:ammonia_static_polarizability}.
We use a (8e, 28o) active space, which contains 419 million determinants.
The structure of ammonia is $C_{3v}$ symmetric, which is represented in practice as the Abelian $C_s$ group in our calculations.
The reference FCI $\alpha_{xx}$ and $\alpha_{yy}$ polarizability components are equal (around 9.407 a.u.), while the $\alpha_{zz}$ component is slightly smaller (around 6.568 a.u.). 
Overall, we find that the convergence behavior is quite similar to that of water. 
The GS+V and GS+V+X schemes, which include a coupling to the one-electron operator, converge well for all three non-zero components of the polarizability. The GS and GS+X schemes converge to the correct limiting value for the $\alpha_{zz}$ component, but fail at describing the $\alpha_{xx}$ and $\alpha_{yy}$ components, converging to two other limiting values (2.35 a.u. and 7.05 a.u.), which summed together is equal to $\alpha_{xx}$ or $\alpha_{yy}$.
Again, this behaviour can be understood from symmetry considerations. 

From the middle panel of Figure \ref{fig:ammonia_static_polarizability}, we see that convergence to an accuracy of 0.1 a.u. requires 36,000 (GS+V+X) or 75,000 (GS+V) determinants, while a higher accuracy of 0.01 a.u. requires 650,000 or 4,000,000 determinants. In comparison to water, an equivalently tight convergence of the polarizability in this larger variational space requires a corresponding larger number of determinants. 

Overall, we find that although all four tested schemes can converge towards the FCI limit in some cases (the $\alpha_{zz}$ components of water and ammonia), including a coupling to the one-electron operator (+V) is essential to achieve robust convergence more generally.
The additional coupling to the response vector (+X) gives modest (but not dramatic) improvements to the convergence rate.

\FloatBarrier
\subsection{Damped response and core-level spectroscopy of the water molecule}

Next, we turn to computing core-level absorption spectra of water.
The absorption strength can be related to the imaginary part of the damped, frequency-dependent complex dipole--dipole polarizability (see Eq. \eqref{eq:sigma_abs}), which can be obtained using (damped) linear response.
In XAS, an electron is excited from the core level into the virtual orbitals. Therefore, a qualitatively correct description requires the inclusion of the relevant core orbitals in the active space.
Predicting XAS using a SCI framework is anticipated to be more challenging than the previous examples of static polarizabilities, since the core absorption spectrum requires a description of highly excited states, which are unlikely to be well described with only the determinants important for the ground state energy.
We first consider a smaller example where obtaining CASCI reference results remains easily accessible, using a (10e, 14o) active space (with around 4 million determinants). We note that this is a very limited active space and basis set, but it is important to verify whether the LR-SCI method converges to the correct FCI (or CASCI) limit, regardless of any specific choice of active space and basis.
We compute spectra with LR-SCI with a set of increasingly tight thresholds $\varepsilon$.

\begin{figure}
    \centering
    \includegraphics[width=0.8\linewidth]{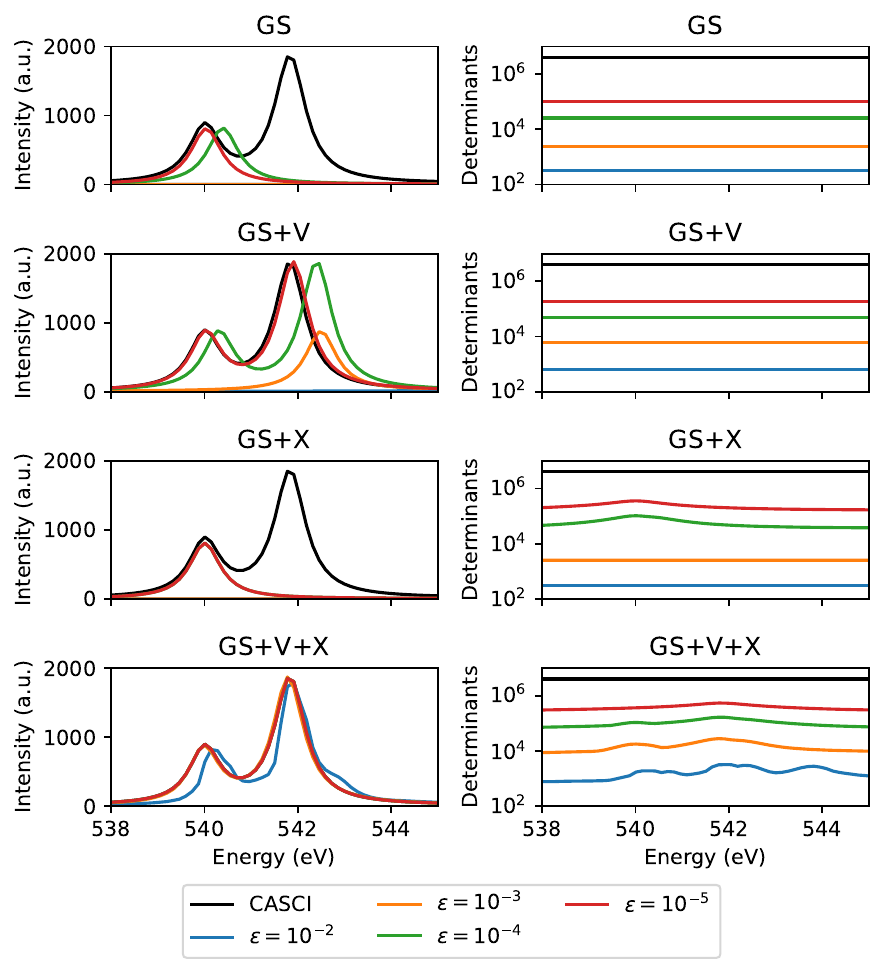}
    \caption{Water/cc-pVDZ K-edge X-ray absorption spectrum. A (10e, 14o) active space is used. The spectrum is computed from the isotropically averaged imaginary part of damped polarizability, with $\gamma=0.4$ eV. The damped polarizability is computed with a 0.005 a.u. frequency resolution. The left panels show the computed absorption spectrum, while the right panels show the number of determinants included in the CI expansion (log scale) at a given threshold $\varepsilon$. Different SCI coupling models are tested (see panel titles). CASCI (black lines) results serve as the reference result.}
    \label{fig:water_xas_calibration}
\end{figure}

As shown in Figure~\ref{fig:water_xas_calibration}, with the (10e, 14o) active space, there are two primary transitions in the XA spectrum of water, namely a lower-intensity absorption at 540.0 eV, followed by a more intense absorption at 541.7 eV, which correspond to the 4$a_1$ and 2$b_2$ transitions\cite{schirmer1993k,ekstrom2006x,fahleson2016polarization}. A third, higher-energy Rydberg-like transition (2$b_1$) is present experimentally, but a theoretical description would require the inclusion of diffuse basis functions, which are missing in the (10e, 14o) space.

From the top panel of Figure \ref{fig:water_xas_calibration}, we see that using only the ground-state HCI screening for adding selecting determinants fails to reproduce the spectrum, even with relatively large determinant spaces.
By the time about 120,000 determinants are included (red lines), the first absorption band at 540 eV is finally reproduced, although with a slightly underestimated intensity. However, the second, more intense absorption at 541.7 eV is missing completely. Similar to the previously discussed static polarizabilities, this behavior can be understood from symmetry arguments, namely that the HCI ground-state selection will only pick out determinants with $A_1$ symmetry, meaning that only $z$-polarized transitions (like 4$a_1$) appear in the absorption spectrum of water.

By adding a coupling to the property vector (GS+V), results can be improved significantly, and good agreement with the CASCI reference is obtained as the threshold $\varepsilon$ is tightened. Interestingly, we find examples where the intensities of the spectrum are quite good even though the peak positions are significantly blue-shifted (see $\varepsilon=10^{-4}$, green lines). For such cases, the most important determinants for both the ground state and any determinants connecting through the dipole operator are included, which ensures that the property vector is quite accurate. However, the response vector is not well captured, resulting in a significant blue shift.
With tighter thresholds, the peak positions eventually become well-described ($\varepsilon=10^{-5}$, red lines).

By including a coupling to the response vector (GS+X), we obtain spectra that look overall quite similar to just including the ground-state HCI determinant selection, with only the first absorption band being reproduced in the spectrum, due to missing important determinants required to represent the property vector. Nevertheless, closer inspection reveals that the transition energy of the lower-energy peak is reproduced decently well, while the intensities are too low, even at the tightest $\varepsilon$. 

With coupling to both the property vector and the response vector (GS+V+X), some of the qualitative behaviour of the spectrum is captured already at very loose thresholds ($\varepsilon=10^{-2}$, blue lines), although the spectrum appears quite noisy and is blue-shifted by about 0.4 eV.
By tightening the thresholds, near-quantitative agreement with the CASCI reference is achieved. At the scale plotted, the CASCI (black lines) and SCI ($\varepsilon = 10^{-3}$, orange lines) overlap almost completely.
Further tightening the thresholds improves agreement, although it is hard to distinguish on the plotted scale.
One should keep in mind that the extra determinant addition steps come at a cost. Thus, we will next consider the convergence characteristics (spectrum error against the number of determinants).

\begin{figure}
    \centering
    \includegraphics[width=0.4\linewidth]{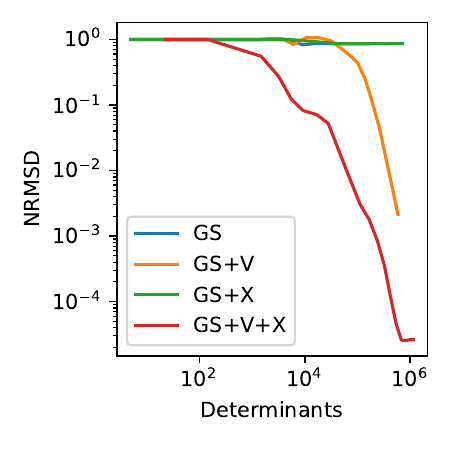}
    \caption{RMSD in the isotropically averaged imaginary part of damped polarizability, with $\gamma=0.4$ eV (see Figure \ref{fig:water_xas_calibration}) against the maximum number of determinants required across the frequency grid.}
    \label{fig:water_xas_convergence}
\end{figure}

Figure \ref{fig:water_xas_convergence} shows the convergence characteristics in further detail, using a denser grid of thresholds $\varepsilon$.
To quantify the convergence, we plot the normalized root mean squared deviation (NRMSD)  in the isotropic part of the imaginary part of the damped polarizability, defined as
\begin{equation}
    \mathrm{NRMSD}=\frac{\sqrt{\frac{1}{N}\sum_{i=1}^{N} (\mathrm{Im}(\bar{\alpha}^{\mathrm{CASCI}}(\omega_i)) -\mathrm{Im}(\bar{\alpha}^{\mathrm{SCI}}(\omega_i)))^2}} {\sqrt{\frac{1}{N}\sum_{i=1}^{N} (\mathrm{Im}(\bar{\alpha}^{\mathrm{CASCI}}(\omega_i))^2}}.
\end{equation}
In terms of spectrum reproduction, lower values of the NMRSD are better. The normalization is selected such that a zero spectrum gives an NMRSD of 1.
We find that even though the more elaborate GS+V+X scheme adds more determinants for a given threshold $\varepsilon$, the overall ``determinant economy'' is favorable. With GS+V+X (red lines), we find fast convergence in the spectrum with the number of determinants. An NRMSD below 0.01 is achieved with about 47,000 determinants, which is significantly more compact than the 390,000 determinants required by GS+V (orange lines). The two remaining schemes, GS and GS+X, are not able to achieve any good convergence in the spectrum, which was already clear from the many missed transitions from Figure \ref{fig:water_xas_calibration}. 

The (10e, 14o) active space considered so far is well within the reach of a CASCI expansion, but we now turn to calculation in larger orbital spaces, for which FCI is no longer practical.
Based on the numerical results from Figures \ref{fig:water_xas_calibration} and \ref{fig:water_xas_convergence}, we will use the GS+V+X scheme in the following. The smaller test system suggested that good results could be obtained with $\varepsilon=10^{-4}$. Additional calculations were also performed with the tighter threshold $\varepsilon=5\times 10^{-5}$.
We carry out calculations with the cc-pVDZ basis set in the full (10e, 24o) space, with aug-cc-pVDZ (10e, 41o), and with d-aug-cc-pVDZ (10e, 58o).
We also include experimental results of the NEXAFS spectrum of gas-phase water from Ref. \citenum{schirmer1993k} (black lines). 
The computed spectra are rigidly shifted to align with the experimental 4$a_1$ transition.

\begin{figure}
    \centering
    \includegraphics[width=0.55\linewidth]{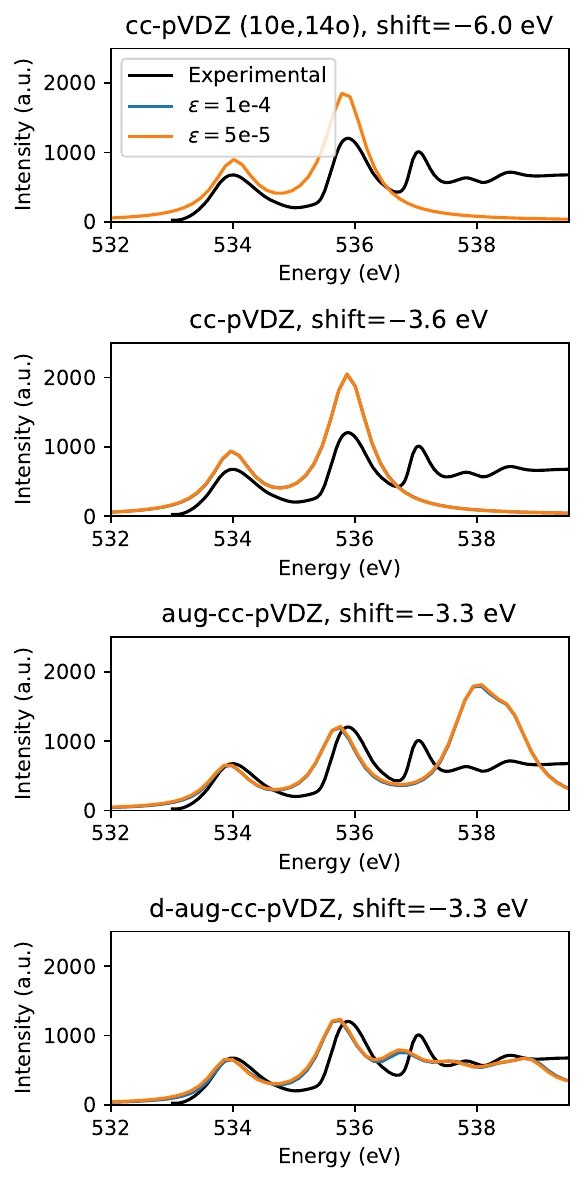}
    \caption{Water K-edge X-ray absorption spectra computed with the GS+V+X model (blue and orange lines). In the top panel, an active space is used, while the remaining panels are full-space calculations. Increasing basis set expansions (from cc-pVDZ to d-aug-cc-pVDZ) are adopted. The experimental spectrum is shown in black. The computed spectra are shifted to align with the first transition of the experimental spectrum.}
    \label{fig:water_xas_big}
\end{figure}

From Figure \ref{fig:water_xas_big}, we see that going to larger basis set expansions significantly impacts the computed spectrum.
First, going from the smaller (10e, 14o) space to the full (10e, 24o) space with cc-pVDZ red-shifts the spectrum by about 2.4 eV, but otherwise keeps the features of the spectrum mostly intact. 
Comparing with the experimental gas-phase absorption spectrum of water, we find passable agreement on the position and splitting of the first two absorption features. However, the third Rydberg-like transition is completely missing.
We find almost no discernible difference between the spectra computed with $\varepsilon=10^{-4}$ and $\varepsilon=5\times10^{-5}$, suggesting that a reasonable convergence towards the FCI result has been obtained in this basis.

Upon adding augmented functions (aug-cc-pVDZ), the spectrum features three absorption bands, with the first two features aligning better with the experimental spectrum. 
The computed spectrum also requires a smaller shift to align with the experimental spectrum ($-$3.3 eV) than with cc-pVDZ ($-$3.6 eV).
The Rydberg-like transition is finally present, but occurs at too high an energy and exhibits an artificially high absorption strength.
In the larger aug-cc-pVDZ basis, there is reasonably good agreement between the spectra computed with $\varepsilon=10^{-4}$ and $\varepsilon=5\times10^{-5}$, although there are a few noisy spikes in the spectrum computed with the looser threshold.

Upon going to the d-aug-cc-pVDZ basis, agreement with the experimental spectrum further improves. 
The position and intensity of the first two absorption bands (4$a_1$ and 2$b_2$) align quite well with the experimental gas-phase spectrum. The description of the Rydberg-like transition is improved, appearing closer and with an intensity more similar to the experimental spectrum. However, the agreement is clearly still not perfect, with the transition appearing at slightly too low energies and with too low intensity.
As before, there is reasonably good agreement between the spectra computed with $\varepsilon=10^{-4}$ and $\varepsilon=5\times10^{-5}$.
It is noteworthy that with the d-aug-cc-pVDZ basis, a large shift of $-$3.3 eV is still required to align the experimental and computed spectra. The required shift could likely be reduced by further increasing the basis set size and by including an account of relativistic effects.

\FloatBarrier
\subsection{NMR spin-spin coupling constants}
Next, we evaluate the performance of LR-SCI for computing nuclear spin-spin coupling constants. These are demanding test cases because they depend significantly on electron correlation effects and arise from four distinct physical contributions \cite{ramsey_electron_1953,helgaker2008quantum,spasB8}: the diamagnetic spin-orbit (DSO), paramagnetic spin-orbit (PSO), Fermi contact (FC), and spin-dipolar (SD) terms. The DSO contribution can be evaluated as an expectation value of the ground-state wave function.
The remaining three terms require linear response calculations: the PSO term is obtained from an imaginary singlet perturbation, while the FC and SD terms originate from real triplet perturbations. 
Our linear response implementation is purely determinant-based and does not take advantage of spin-adaptation, so the distinction between singlet/triplet response does not require any special considerations.
The calculation of spin-spin coupling constants has slightly atypical basis set requirements, which can be addressed by using special, property-optimized basis sets\cite{jensen2006basis,jensen2010optimum,benedikt2008optimization,kjaer2011pople}. These include tight $s$-functions, which are essential for the Fermi contact term, as this term depends on the electron density on the nucleus. 
For the same reason, it is typically necessary to include all core electrons in the active space, as they contribute substantially to the Fermi contact term.
Apart from this, tight $p$-, $d$- and $f$-functions improve the spin-dipole and paramagnetic spin-orbit contributions.

\begin{figure}
    \centering
    \includegraphics[width=0.7\linewidth]{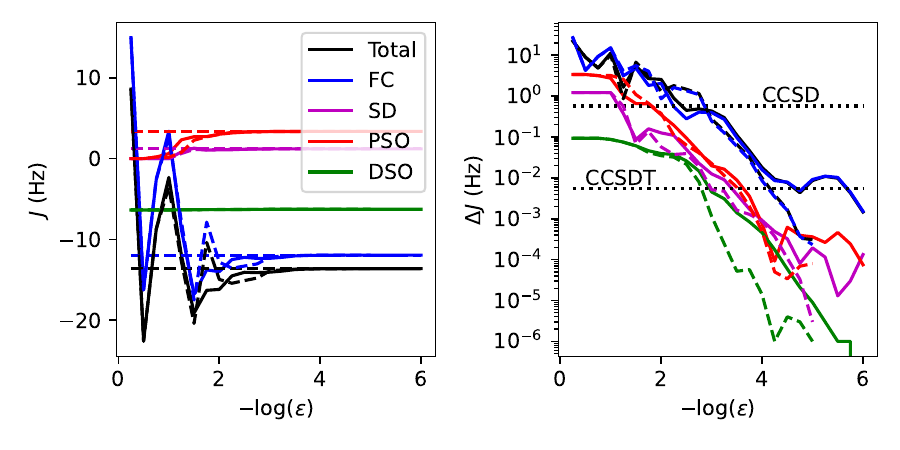}
    \caption{Spin-spin coupling constant $^2J_{\mathrm{HH}}$ of the water molecule with a 6-31G-J basis. The left panel plots the computed coupling constant as a function of the threshold $\varepsilon$ (left) on a linear scale. The dashed lines indicate the FCI result. The error in $^2J_{\mathrm{HH}}$ relative to the FCI reference is shown in the right panel (log scale). The dotted lines in the right panels indicate the accuracy of CCSD and CCSDT in the total $^2J_{\mathrm{HH}}$. Solid lines indicate results obtained using the canonical HF basis, while dashed lines indicate the use of a natural orbital basis.}
    \label{fig:sscc_H2O_6-31G-J}
\end{figure}

Figure \ref{fig:sscc_H2O_6-31G-J} shows the convergence of the $^2J_{\mathrm{HH}}$ spin-spin coupling constant computed with LR-SCI (GS+V+X) of the water molecule with a 6-31G-J basis\cite{kjaer2011pople}. 
Calculations are carried out in the full orbital space (10e, 20o), which is small enough that obtaining reference FCI values is feasible (with which we obtain a coupling constant of $-13.62$ Hz). 
All four terms contribute significantly to the spin-spin coupling constant, with the largest in magnitude being the FC term ($-11.94$ Hz), followed by the DSO term ($-6.27$ Hz), the PSO term ($3.37$ Hz), and the SD term ($1.21$ Hz).
We find that LR-SCI with the GS+V+X coupling scheme converges systematically towards the FCI limit as $\varepsilon$ is decreased.
The convergence is limited by the FC term, which requires a tighter $\varepsilon$ than the two remaining response properties for the same accuracy.

The difficulty in calculating the FC term is further amplified by higher determinant requirements than the PSO and SD terms at a given $\varepsilon$, as shown in Table \ref{tab:sscc_H2O_6-31G-J_data}. 
For example, with $\varepsilon=10^{-6}$, calculating the FC term requires a 7.2-fold increase in the number of determinants over the ground-state HCI, compared to just 2.5 and 1.7-fold increases for the PSO and SD terms.
The FC operator depends on the electronic wave function amplitude at the position of a nucleus, leading to matrix elements that are numerically much larger and more widely distributed than for the PSO and SD operators.
As a consequence, the FC term becomes computationally challenging since the property and response vectors exhibit less sparsity than those associated with the PSO and SD operators (see Figure \ref{fig:si_vector_structure}).
Thus, the FC term constitutes the primary bottleneck for achieving balanced convergence among the SSCC contributions.

We carried out calculations in both the canonical HF basis and in the natural orbital (NO) basis\cite{lowdin1955quantum} (Figure \ref{fig:sscc_H2O_6-31G-J}, dashed lines). 
NOs are the set of orbitals which diagonalize the one-electron density matrix, and are a popular choice in (S)CI calculations\cite{abrams2004natural,illas1988approximate,loos2018mountaineering}, since they are known to speed up the convergence towards the FCI limit for energies.
The calculations in the NO basis do indeed show slightly faster convergence in the spin-spin coupling constant than the calculations in the canonical HF basis.
We note that using NOs as the orbital basis is likely not an optimal choice in terms of describing response properties. While NOs improve the description of the ground state, which in turn will lead to better accuracy in the following linear-response related quantities, NOs are probably not the most compact orbital basis for response properties. 
As shown by \citeauthor{kumar2017frozen}, truncation in the NO space leads to higher errors in the polarizability compared to truncation in the canonical HF orbitals\cite{kumar2017frozen}.
Ideally, the orbitals should give a balanced description of both the ground state and the response vectors.
Thus, it would be interesting to explore, e.g., response-aware orbital sets \cite{crawford2019reduced,d2020pno++}.

\begin{table}
    \centering
\begin{tabular}{|c|rrrr|rrrr|}
\hline 
$\varepsilon$ & $N_{\mathrm{det}}^{\mathrm{GS}}$ & $N_{\mathrm{det}}^{\mathrm{PSO}}$ & $N_{\mathrm{det}}^{\mathrm{FC}}$ & $N_{\mathrm{det}}^{\mathrm{SD}}$ & $^{2}J_{\mathrm{HH}}^{\mathrm{DSO}}$ & $^{2}J_{\mathrm{HH}}^{\mathrm{PSO}}$ & $^{2}J_{\mathrm{HH}}^{\mathrm{FC}}$  & $^{2}J_{\mathrm{HH}}^{\mathrm{SD}}$\tabularnewline
\hline 
$10^{-3}$ & 9,110 & 18,258 & 134,515 & 14,598 & -6.27103 & 3.35373 & -12.33865 & 1.20233\tabularnewline
$10^{-4}$ & 97,249 & 203,289 & 1,137,959 & 146,379 & -6.26695 & 3.37276 & -11.95822 & 1.21395\tabularnewline
$10^{-5}$ & 603,516 & 1,398,888 & 5,475,649 & 927,652 & -6.26650 & 3.37388 & -11.95157 & 1.21506\tabularnewline
$10^{-6}$ & 2,427,520 & 6,083,774 & 17,590,596 & 4,102,234 & -6.26649 & 3.37344 & -11.94393 & 1.21500\tabularnewline
\hline 
CCSD &  &  &  &  & -6.26973 & 3.37847 & -12.51860 & 1.21731\tabularnewline
CCSDT &  &  &  &  & -6.26802 & 3.37438 & -11.94770 & 1.21535\tabularnewline
\hline 
FCI & 240,374,016 &  &  &  & -6.26649 & 3.37352 & -11.94241 & 1.21486\tabularnewline
\hline 
\end{tabular}
    \caption{Individual contributions to the $^2J_{\mathrm{HH}}$ spin-spin coupling constant (in Hz) for water/6-31G-J using SCI, CCSD, and CCSDT, with FCI providing reference results. The maximum number of determinants required in the GS+V+X linear response calculation is also reported.}
    \label{tab:sscc_H2O_6-31G-J_data}
\end{table}

From Figure \ref{fig:sscc_H2O_6-31G-J}, we see that an error below 0.1 Hz in the total coupling constant is obtained after $\varepsilon=10^{-3.75}$ (or $10^{-3.50}$ with NOs), while an error below 0.005 Hz can be obtained with $\varepsilon$ below $10^{-5.75}$ (or $10^{-4.25}$ with NOs).
For comparison, we also carried out coupled-cluster calculations, which yield errors of 0.57 Hz (CCSD) or 0.005 Hz (CCSDT) relative to the FCI reference.
We are thus able to obtain spin-spin coupling constants beyond the accuracy of standard wave-function-based methods such as CCSDT (at least when using small basis sets), but it is also clear that the accuracy of CCSDT for weakly correleated, closed-shell systems, would likely be sufficient for most practical purposes.

In Table \ref{tab:sscc_H2O_larger_basis}, we report the $^2J_\mathrm{HH}$ coupling constant in a series of larger basis sets, namely 6-31+G*-J (29 orbitals), 6-31++G**-J (37 orbitals), and 6-311++G**-J (46 orbitals). 
We conducted LR-SCI (GS+V+X) calculations with evenly logarithmically spaced thresholds $\varepsilon$, spaced 0.25 log-units apart.  We report results from the best calculation that we could manage to complete within computational constraints.
We have included an error estimate for the LR-SCI estimated value based on the magnitude of the change from the next-best $\varepsilon$.
With the computational resources at our disposal (128-core nodes with up to 4TB of memory), we managed LR-SCI calculations with up to $1.5\times10^8$ determinants. 
Larger calculations could likely be achieved through a combination of further code optimization and hardware advances.
PyCI computes and stores Hamiltonian matrix elements in a (lower-triangular) compressed sparse row matrix format, which, for large CI expansions, will require a significant amount of memory to store. For the systems considered here, the sparse matrix typically stores around 1000 non-zero elements per row (although this figure is system-dependent), which for $1.5\times10^{8}$ determinants leads to a memory footprint of around 2.4TB for storing the indices and data. Thus, significantly larger determinant expansions are out of reach for the present implementation and would require a direct matrix-vector multiplication routine or multinode distributed sparse matrix storage. 
A detailed cost breakdown of the individual steps of the GS+V+X LR-SCI calculations required for the calculation of the spin-spin coupling constant is shown in Table \ref{tab:si_sscc_detailed_timings}, from which we find that the computational effort is mainly dominated by steps related to computing the FC term. 
The additional computational effort can mostly be attributed to increased determinant requirements, and the required number of Hamiltonian matrix-vector products is not significantly increased compared to the PSO and SD operators (see Table \ref{tab:si_sigma_vectors}).

From Table \ref{tab:sscc_H2O_larger_basis}, we see that the LR-SCI prediction of the spin-spin coupling constant of water generally agrees very well with the CCSDT-computed values.
The spin-spin coupling constant gradually decreases in magnitude as the size of the basis set increases, going from $-$13.6 Hz (with 6-31G-J) to about $-$10.1 Hz (with 6-311++G**-J).
\citeauthor{faber2017importance} has reported\cite{faber2017importance} CCSDT calculations on the same system with an even larger aug-ccJ-pVTZ basis, yielding a $^2J_\mathrm{HH}$ of $-$7.79 Hz, which suggests that the basis set description is not completely saturated in the present set of calculations.
Up to and including the calculation with 6-31++G**-J, the LR-SCI error estimate remains low (below $6\times10^{-3}$ Hz), and we can, with reasonable confidence, suggest that the CCSDT-computed values are within 0.01 Hz of the FCI limit.
For the largest 6-311++G**-J basis, the LR-SCI error estimate is on the same order of magnitude as the difference between the CCSDT and LR-SCI-computed values for the coupling constant.
The calculations on the smaller basis sets suggest a rather consistent agreement between the LR-SCI and results for $^2J_\mathrm{HH}$, so the more substantial difference with the largest 6-311++G**-J basis is almost certainly due to a not completely converged (in $\varepsilon$) LR-SCI calculation.
Again, we find that the calculations in the NO basis show slightly faster convergence for the spin-spin coupling constant than the calculations in the canonical HF basis.
As a result, we find improved agreement with the CCSDT result, reducing the difference between our best LR-SCI (GS+V+X, NOs) and the CCSDT calculation to -0.005 Hz, which is in line with the deviations from the smaller basis sets.

\begin{table}
    \centering
\begin{tabular}{|c|cccc|cc|cc|}
\hline 
Basis & $\varepsilon$ & $N_{\mathrm{det}}^{\mathrm{max}}$ & $^{2}J_{\mathrm{HH}}^{\mathrm{SCI}}$ & Err.est. & $^{2}J_{\mathrm{HH}}^{\mathrm{CCSD}}$ & $\Delta^{2}J_{\mathrm{HH}}^{\mathrm{CCSD}}$ & $^{2}J_{\mathrm{HH}}^{\mathrm{CCSDT}}$ & $\Delta^{2}J_{\mathrm{HH}}^{\mathrm{CCSDT}}$\tabularnewline
\hline 
6-31G-J & $10^{-6.00}$ & 17,590,596 & -13.622 & 0.0028 & -14.192 & -0.570 & -13.626 & {-0.004}\tabularnewline
6-31+G{*}-J & $10^{-5.75}$ & 159,508,181 & -12.263 & 0.0011 & -12.759 & -0.496 & -12.274 & {-0.011}\tabularnewline
6-31++G{*}{*}-J & $10^{-5.25}$ & 131,567,679 & -10.576 & 0.0060 & -11.041 & -0.465 & -10.583 & {-0.007}\tabularnewline
6-311++G{*}{*}-J & $10^{-4.75}$ & 125,363,640 & -10.167 & 0.0435 & -10.644 & -0.477 & -10.137 & {\phantom{-}0.030}\tabularnewline
\hline 
6-311++G{*}{*}-J$^{a}$ & $10^{-4.75}$ & 81,318,131 & -10.132 & 0.0115 &  & -0.512 &  & {-0.005}\tabularnewline
\hline 
\end{tabular}    \caption{$^2J_\mathrm{HH}$ (in Hz) spin-spin coupling constants computed with LR-SCI (GS+V+X), CCSD, and CCSDT. For the SCI calculations, the best value of $\varepsilon$ used and the maximum number of determinants required are also reported. The error estimate is based on the magnitude of the change in $^2J_\mathrm{HH}$ from the next-best $\varepsilon$. $^a$using a natural orbital basis for the LR-SCI.}
    \label{tab:sscc_H2O_larger_basis}
\end{table}

For systems such as the water molecule, the accuracy provided by triples-including coupled-cluster methods is sufficient for most purposes.
At the same time, the computational efficiency of traditional coupled-cluster methods is, at least for the present implementation, also superior to LR-SCI (see Table \ref{tab:si_timings_vs_cc}). 
Nevertheless, our LR-SCI implementation can validate the performance of electron-structure methods beyond the limits of where exact FCI is feasible, and may be useful in cases where coupled-cluster methods fail.
CCSDT is expected to be reliable for chemical systems that do not exhibit strong static correlation, but there remain cases with pronounced multireference character where additional validation is worthwhile. As shown in Table \ref{tab:si_polarizability_extra_molecules}, some supposedly multiconfigurational systems (like the carbon dimer) turn out to be unproblematic. Other systems, like the boron nitride, reveal more substantial errors for coupled-cluster-based methods. Finally, one can consider (slightly contrived) systems, like the N$_2$ molecule at stretched bond lengths, where CCSDT fails to give even a qualitatively correct description. 

Looking ahead, several directions could enhance the accuracy, stability, and scope of LR-SCI beyond the relatively small examples considered so far. 
First and foremost, incorporating perturbative corrections, analogous to the second-order treatments commonly used for SCI energies, could be a systematic route to recover residual contributions from the external space, extending applicability to larger systems. 
Adopting a spin-adapted CSF basis would reduce the number of independent amplitudes compared to a determinant representation, which might improve compactness and numerical robustness. 
Preliminary tests indicate that natural orbitals accelerate convergence relative to canonical HF orbitals. However, as discussed above, there is a tension between orbitals that are optimal for describing the ground state and those that best capture response properties. 
More sophisticated orbital choices that balance these requirements, or explicit orbital optimization, may therefore provide additional efficiency gains. 
Collectively, these developments could bring LR-SCI closer to a robust, systematically improvable approach for response theory suitable for larger systems.

\section{Conclusion}\label{sec:conclusion}
In this work, we have formulated linear response theory for SCI wave functions and demonstrated its applicability to a range of molecular properties. While SCI has so far primarily been used for ground- and excited-state energies, our work shows that response properties can also be systematically converged toward the FCI limit by augmenting the determinant selection procedure. In particular, we have explored two response-theory motivated selection criteria, which extend the ground-state SCI expansion to include determinants relevant for property gradients and response vectors.
While each criterion individually shows good convergence characteristics for certain cases, we find that only their combined use (GS+V+X) provides consistently robust convergence across all tested properties.

We have conducted benchmark calculations of static polarizabilities of water and ammonia, demonstrating that LR-SCI can recover reference FCI static polarizabilities with high accuracy. 
Using a damped linear response framework, we have computed the K-edge X-ray absorption spectrum of water and found good agreement with reference CASCI spectra in smaller active spaces. 
Using a full-space treatment, we find that the agreement with experimental gas-phase X-ray absorption spectra improves significantly as the basis set is expanded, with good qualitative agreement obtained with the d-aug-cc-pVDZ basis.
Finally, calculations of NMR spin-spin coupling constants showed that all four physical contributions can be treated within (LR-)SCI, with convergence being limited mainly by the Fermi contact term. For small basis sets, where reference FCI results could be obtained, we found that LR-SCI can reach accuracies beyond CCSDT.

Altogether, these results demonstrate that LR-SCI provides a systematically improvable framework for molecular response properties with near-FCI accuracy. 
Practical applications are still limited by determinant-space growth and memory requirements, which means that we have only been able to carry out high-accuracy calculations on relatively small molecules.
Even so, the method already enables high-level benchmarks for linear response properties. 
With further algorithmic advances, LR-SCI has the potential to extend the reach of accurate response theory to larger systems and more complex spectroscopic observables.

\begin{acknowledgement}
Computations/simulations for the work described herein were supported by the DeIC National HPC Centre, SDU.
We acknowledge the financial support of the Novo Nordisk Foundation for the focused research project \textit{Hybrid Quantum Chemistry on Hybrid Quantum Computers} (HQC)$^2$, grant number NNFSA220080996.
\end{acknowledgement}

%\begin{suppinfo}
%...
%\end{suppinfo}

%\section*{Data Availability Statement}
%TBD

\bibliography{main}

\end{document}

\section{Molecular Geometries}
\begin{lstlisting}
3
water
  O          -0.00000000000000      0.00000000000000      0.07795473143103
  H           0.75756859835931      0.00000000000000     -0.51001699058892
  H          -0.75756859835931      0.00000000000000     -0.51001699058892
\end{lstlisting}

\begin{lstlisting}
4
ammonia
  N           0.00000000000000      0.00000000000000     -0.06813337282263
  H           0.93730546833930      0.00000000000000      0.31558999772704
  H          -0.46865273416965      0.81173034668791      0.31558999772704
  H          -0.46865273416965     -0.81173034668791      0.31558999772704
\end{lstlisting}

\section{Additional analyses}
\begin{figure}
    \centering
    \includegraphics[width=0.5\linewidth]{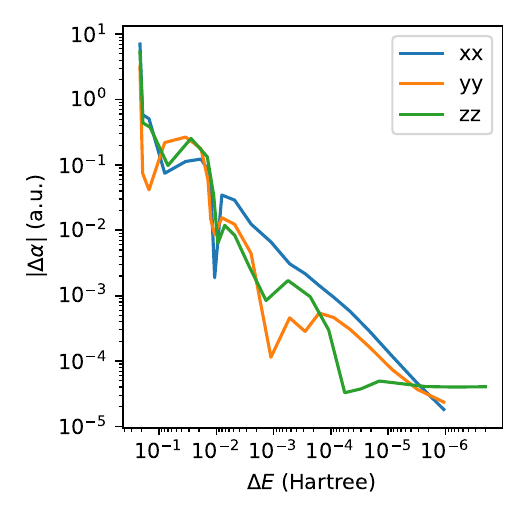}
    \caption{Error in the $xx$, $yy$, and $zz$ polarizability components of water/cc-pVDZ (8e, 23o) from a GS+V+X LR-SCI calculation against the error in the electronic energy, using (frozen-core) FCI as a reference.}
    \label{fig:si_energy_vs_polarizability}
\end{figure}

Figure \ref{fig:si_energy_vs_polarizability} plots the error in the three polarizability components of water from an LR-SCI calculation against the error in the electronic energy, using FCI as a reference.
Of note, a relatively loose energy accuracy of $10^{-3}$ Hartree already yields polarizabilities accurate to better than 0.01 a.u.

\begin{figure}
    \centering
    \includegraphics[width=1.0\linewidth]{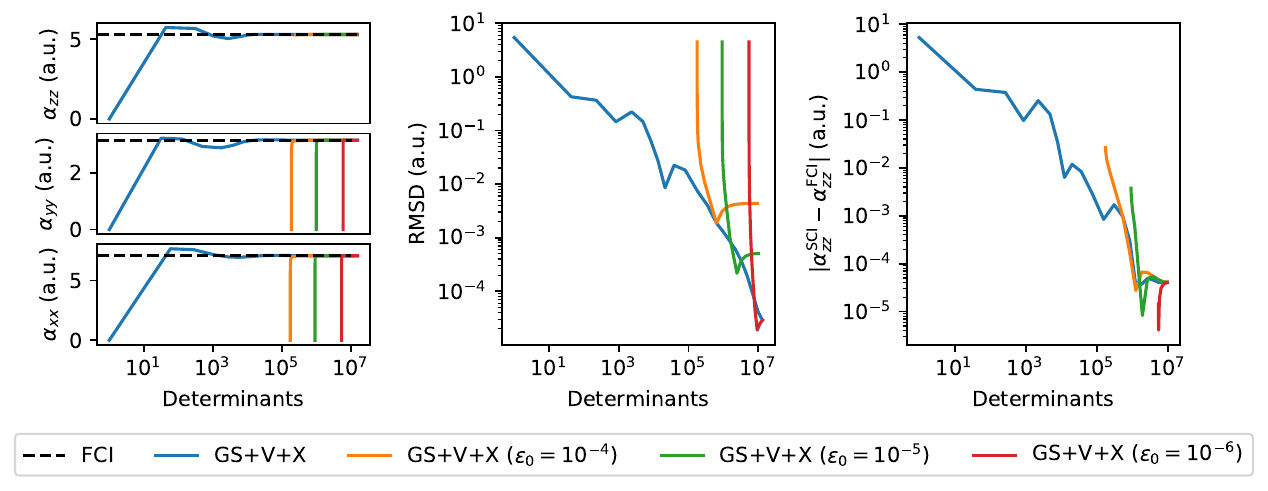}
    \caption{Accuracy in the GS+V+X LR-SCI polarizability of water/cc-pVDZ (8e, 23o) relative to the FCI solution. Results are reported for GS+V+X with variable $\varepsilon_0=\varepsilon_V=\varepsilon_X$, and for a series of calculations where $\varepsilon_0$ is instead fixed at $10^{-4}$, $10^{-5}$, or $10^{-6}$, with $\varepsilon_V=\varepsilon_X$ variable.  
    The left panels show the error in the $xx$, $yy$, and $zz$ components (on a linear scale). The middle panel shows the RMSD in the polarizability (log scale). The right panel shows the error in the $zz$-component of the polarizability (log scale).}
    \label{fig:si_fixed_epsilon}
\end{figure}

In the main manuscript, we generally adopted equal thresholds for the ground-state HCI selection ($\varepsilon_0$), and for the determinant-addition steps related to the property-gradient ($\varepsilon_V$) and the response-vector ($\varepsilon_X$).
One might ask whether it is beneficial to converge the ground-state HCI tightly before attempting any linear response calculations.
To assess whether this approach is beneficial, we include another series of calculations, where $\varepsilon_0$ is fixed to some specific value (e.g., $\varepsilon_0=10^{-6}$), and only $\varepsilon_V=\varepsilon_X$ is variable.
From Figure \ref{fig:si_fixed_epsilon}, it is clear that there must be a balance in the thresholds. If $\varepsilon_0 \ll \varepsilon_V=\varepsilon_X$, the +V and +X steps have almost no effect, and the GS+V+X model approximately reduces to the LR-SCI with only GS selection. 
This is reflected in the high initial errors in the $xx$ and $yy$ components (Figure \ref{fig:si_fixed_epsilon}, middle panel), which are only reduced as the two response-related criteria are tightened (i.e., approaching $\varepsilon_0=\varepsilon_V=\varepsilon_X$). 
On the other hand, if $\varepsilon_V=\varepsilon_X$ is tightened beyond $\varepsilon_0$, the accuracy again suffers, since response-related quantities (like the property gradient) are dependent on the accuracy of the ground-state CI vector.

\begin{figure}
    \centering
    \includegraphics[width=1.0\linewidth]{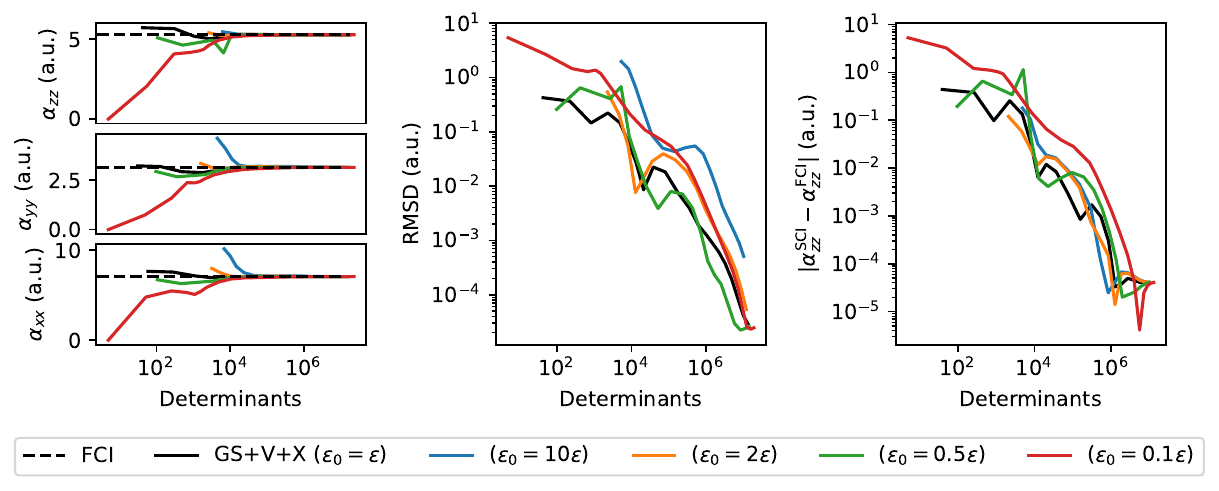}
    \caption{Accuracy in the GS+V+X LR-SCI polarizability of water/cc-pVDZ (8e, 23o) relative to the FCI solution. Results are reported for GS+V+X with variable $\varepsilon_V=\varepsilon_X$, while $\varepsilon_0$ is adjusted to have either a looser (blue and orange lines), or tighter threshold (green and red lines) than the response-related thresholds.  
    The left panels show the error in the $xx$, $yy$, and $zz$ components (on a linear scale). The middle panel shows the RMSD in the polarizability (log scale). The right panel shows the error in the $zz$-component of the polarizability (log scale).}
    \label{fig:si_adjusted_epsilon}
\end{figure}

In Figure \ref{fig:si_adjusted_epsilon}, we carry out similar calculations. Instead of taking a fixed $\varepsilon_0$, we keep it variable, but adjust it to be relatively looser or tighter compared to the response-related thresholds $\varepsilon_V$ and $\varepsilon_X$.
If $\varepsilon_0$ is made significantly tighter ($\varepsilon_0=0.1\varepsilon$, red lines), the polarizability converges from below, and the overall performance is worse than the default choice of equal thresholds.
Letting $\varepsilon_0$ instead be significantly looser ($\varepsilon=10\varepsilon$, blue lines), the polarizability converges from above, and the performance again suffers.
In the middle range of $\varepsilon_0$ scalings, the picture is less clear, and we find that determinant counts for which tighter/looser $\varepsilon_0$ can alternatively be beneficial or detrimental. 
These two extremes can be rationalized in terms of the sum-over-states representation of the polarizability
\begin{equation}
        \left<\left<A;B\right>\right>_{\omega} = -\sum_{k\neq 0}\left(\frac{\left<0\left|\hat{A}\right|k\right>\left<k\left|\hat{B}\right|0\right>}{\omega_k-\omega} +\frac{\left<0\left|\hat{B}\right|k\right>\left<k\left|\hat{A}\right|0\right>}{\omega_k+\omega}\right)~,
\end{equation}
where $\omega_k=E_k-E_0$. For the diagonal elements of the ground-state polarizability, the numerators are always positive, and we shall assume for the sake of argument that the transition moments are relatively accurately described.
With $\varepsilon_0$ tight compared to the response-related $\varepsilon$'s, the ground state energy is described very accurately, but the energies of the excited states are typically too high, meaning that the $\omega_k$ are generally too large, which will underestimate the polarizability.
Similarly, if too much emphasis is put on the response-related $\varepsilon$'s, the ground-state energy is too high, which makes the $\omega_k$ in the denominator too low. As a result, the polarizability becomes overestimated.

\begin{figure}
    \centering
    \includegraphics[width=0.5\linewidth]{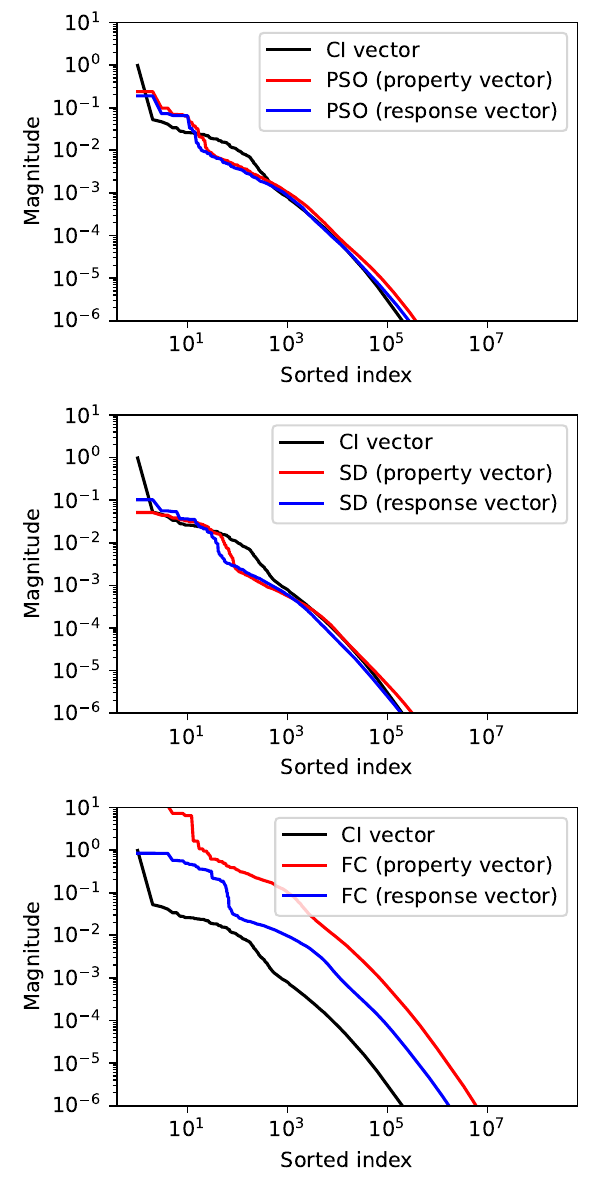}
    \caption{Sorted magnitude plots of the CI vector, and property and response vectors (FCI) for water/6-31G-J. The PSO (H$_1$, $x$), SD (H$_1$, $xx$), and FC (H$_1$) perturbations were selected.}
    \label{fig:si_vector_structure}
\end{figure}

Figure \ref{fig:si_vector_structure} shows the structure of the CI vector, as well as the property and response vectors for a selected PSO, SD, and FC perturbation for water/6-31G-J (FCI). 
The vectors are plotted in terms of the sorted magnitude, which can be used to assess the sparsity of the various vectors.
Notably, the PSO and SD perturbations give vectors, which have a sparsity roughly equivalent to that of the CI vector, while the FC term requires significantly more determinants to describe  (8--25 fold).   

Table \ref{tab:si_polarizability_extra_molecules} shows static polarizabilities computed with LR-SCI compared to results obtained with CCSD and CCSDT for additional molecular systems not considered in the main manuscript.
For the carbon dimer, we used a geometry optimized using CCSD(T)/aug-cc-pCVQZ.
For BN, we used the geometry from Ref. \citenum{karton2017w4}, ($R_{\mathrm{BN}}=1.283$).
For the carbon dimer, which is supposedly a system with pronounced multireference character, the coupled-cluster performs well, with relative errors below 3\% (CCSD) or 1\% (CCSDT).
The BN molecule proves more challenging, with CCSD showing errors of around 17\% in the $\alpha_{xx}$ component, and 11\% in the $\alpha_{zz}$ component. Even CCSDT shows a large error in the $\alpha_{xx}$ component (5\%), but agrees well with the LR-SCI reference for the $\alpha_{zz}$ component, with an error below 0.3\%.
Finally, we consider the N$_2$ molecule at a stretched geometry (i.e., breaking the triple bond), where both CCSD and CCSDT fail at even a qualitatively correct description. The $\alpha_{xx}$ and $\alpha_{yy}$ polarizability components become unequal, and with CCSDT we even find a \emph{negative} $\alpha_{yy}$ component (ground-state polarizabilities should be strictly positive).
The $\alpha_{zz}$ component shows reasonable agreement with LR-SCI, with errors of around 2.5\% (CCSD) and 5.3\% (CCSDT).

\begin{table}
    \centering
\begin{tabular}{|r|r|r|r|rrr|}
\hline 
\multicolumn{2}{|c|}{Molecule} & CCSD & CCSDT & LR-SCI & $\varepsilon$ & \multicolumn{1}{c|}{Err. est.}\tabularnewline
\hline 
\multirow{2}{*}{C$_{2}$/cc-pVDZ} & $\alpha_{xx}=\alpha_{yy}$ & 13.0024 & 12.6488 & 12.7654 & \multirow{2}{*}{$10^{-6.00}$} & \multirow{2}{*}{$5\times10^{-4}$}\tabularnewline
 & $\alpha_{zz}$  & 23.3038 & 23.5516 & 23.6564 &  & \tabularnewline
\hline 
\multirow{2}{*}{C$_{2}$/aug-cc-pVDZ} & $\alpha_{xx}=\alpha_{yy}$ & 19.8897 & 20.3384 & 20.5333 & \multirow{2}{*}{$10^{-5.25}$} & \multirow{2}{*}{$5\times10^{-2}$}\tabularnewline
 & $\alpha_{zz}$  & 25.7869 & 26.1684 & 26.3263 &  & \tabularnewline
\hline 
\multirow{2}{*}{BN/cc-pVDZ} & $\alpha_{xx}=\alpha_{yy}$ & 29.3884 & 23.9635 & 25.1845 & \multirow{2}{*}{$10^{-5.50}$} & \multirow{2}{*}{$3\times10^{-2}$}\tabularnewline
 & $\alpha_{zz}$  & 33.3960 & 30.0165 & 30.1034 &  & \tabularnewline
\hline 
\multirow{3}{*}{N$_{2}$ ($R=2.2$\AA)} & $\alpha_{xx}$ & 5.6416 & 4.9510 & 7.1976 & \multirow{3}{*}{$10^{-5.50}$} & \multirow{3}{*}{$2\times10^{-3}$}\tabularnewline
 & $\alpha_{yy}$ & 6.7717 & $-$21.7613 & 7.1976 &  & \tabularnewline
 & $\alpha_{zz}$ & 17.5893 & 17.0887 & 18.0418 &  & \tabularnewline
\hline 
\end{tabular}
    \caption{Static polarizabilities of selected molecules, computed with CCSD, CCSDT, or GS+V+X LR-SCI. All calculations use the frozen-core approximation (for 1s electrons).}
    \label{tab:si_polarizability_extra_molecules}
\end{table}

\FloatBarrier
\section{Computational performance of LR-SCI}
Table \ref{tab:si_sscc_detailed_timings} shows a timing breakdown of the individual steps involved in calculating the $^2J_\mathrm{HH}$ spin-spin coupling constant of water/6-311++G{*}{*}-J ($\varepsilon=10^{-4.75}$) using GS+V+X LR-SCI.
The PSO and SD terms have relatively modest increases in cost over the ground-state (2-3 fold in the execution time), while the FC term (which has higher determinant requirements) requires significantly larger computational effort.
For the FC term, roughly half the time is spent in routines dominated by computing Hamiltonian matrix-vector products (solving for the CI vector or solving the LR equations). 

For the same system, Table \ref{tab:si_sigma_vectors} reports the number of $\sigma$-vectors (Hamiltonian matrix-vector multiplies) and determinants used in each step during the solution of the linear response equations. Notably, the +V step and the first +X step require the largest amount of $\sigma$-vectors. The subsequent +X steps require fewer $\sigma$-vectors, since the initial guess (a zero-padded solution vector from the previous computational step) is of high quality.

\begin{table}
    \centering
\begin{tabular}{|c|c|c|ccc|}
\hline 
\multicolumn{1}{|c}{Ground state} &  & Response & PSO & SD & FC\tabularnewline
\cline{1-2}\cline{4-6}
$N_{\mathrm{det}}$ & 7,202,373 &  & 17,665,976 & 9,771,321 & 125,363,640\tabularnewline
\hline 
HF & 3 & Misc. & 487 & 490 & 492\tabularnewline
$\left|H_{IJ}c_{J}\right|>\varepsilon_{0}$ & 166 & $\left|B_{IJ}c_{J}\right|>\varepsilon_{V}$ & 27 & 26 & 45\tabularnewline
Build $\mathbf{H}$ & 269 & Build $\mathbf{H}$ & 84 & 15 & 8,121\tabularnewline
Solve $\mathbf{Hc}=E\mathbf{c}$ & 278 & Solve $\mathbf{Hc}=E\mathbf{c}$ & 31 & 30 & 2,626\tabularnewline
1RDM & 275 & Form $\tilde{\mathbf{B}}$ & 245 & 281 & 1,534\tabularnewline
 &  & Solve LR & 83 & 59 & 2,849\tabularnewline
\cline{3-6}
 &  & $\left|H_{IJ}X_{J}\right|>\varepsilon_{X}$ & 152 & 88 & 1,092\tabularnewline
 &  & Build $\mathbf{H}$ & 823 & 145 & 9,115\tabularnewline
 &  & Solve $\mathbf{Hc}=E\mathbf{c}$ & 175 & 117 & 15,630\tabularnewline
 &  & Form $\tilde{\mathbf{B}}$ & 922 & 883 & 9,546\tabularnewline
 &  & Solve LR & 254 & 103 & 10,099\tabularnewline
\hline 
Subtotal & 991 & Subtotal & 3,283 & 2,236 & 61,150\tabularnewline
\hline 
\end{tabular}
    \caption{Timings (in seconds) for various steps involved in a GS+V+X LR-SCI calculation of the $^2J_{\mathrm{HH}}$ spin-spin coupling constant, using water/6-311++G{*}{*}-J ($\varepsilon=10^{-4.75}$) as an example. The response calculation is grouped into parts relating to the +V steps (top) and the +X step (bottom). The values reported for PSO and SD are averages over the 3 or 9 required linear response calculations.}
    \label{tab:si_sscc_detailed_timings}
\end{table}

\begin{table}
    \centering
\resizebox{1.1\textwidth}{!}{%
\begin{tabular}{|c|c|c|c|c|c|c|c|c|c|c|c|c|c|}
\hline 
$\sigma$-vectors & PSO $x$ & PSO $y$ & PSO $z$ & SD $xx$ & SD $xy$ & SD $xz$ & SD $yx$ & SD $yy$ & SD $yz$  & SD $zx$ & SD $zy$ & SD $zz$ & FC\tabularnewline
\hline 
+V & 8 & 10 & 9 & 10 & 8 & 10 & 8 & 9 & 7 & 9 & 5 & 9 & 13\tabularnewline
+X (1) & 8 & 6 & 9 & 7 & 6 & 6 & 8 & 5 & 5 & 6 & 8 & 5 & 9\tabularnewline
+X (2) & 3 & 3 & 3 & 2 & 3 & 2 & 2 & 2 & 2 & 2 & 3 & 2 & 4\tabularnewline
+X (3) & 3 & 2 & 2 & 1 & 2 & 1 & 3 &  - & 2 & 1 & 3 &  - & 3\tabularnewline
\hline 
$N_{\mathrm{det}}$ & PSO $x$ & PSO $y$ & PSO $z$ & SD $xx$ & SD $xy$ & SD $xz$ & SD $yx$ & SD $yy$ & SD $yz$  & SD $zx$ & SD $zy$ & SD $zz$ & FC\tabularnewline
\hline 
+V & 8,365,737 & 8,834,105 & 8,683,993 & 7,425,603 & 7,441,409 & 7,500,741 & 7,441,409 & 7,417,173 & 7,403,407 & 7,500,741 & 7,403,407 & 7,362,499 & 64,326,017\tabularnewline
+X (1) & 16,360,717 & 16,165,659 & 18,403,817 & 10,039,549 & 9,312,157 & 10,947,253 & 9,312,157 & 9,609,897 & 8,767,899 & 10,947,253 & 8,767,899 & 8,759,579 & 123,265,286\tabularnewline
+X (2) & 17,105,531 & 16,443,011 & 19,420,681 & 10,157,305 & 9,586,767 & 11,082,129 & 9,586,767 & 9,694,453 & 8,960,523 & 11,082,129 & 8,960,523 & 8,821,317 & 125,341,690\tabularnewline
+X (3) & 17,118,391 & 16,444,673 & 19,434,865 & 10,158,113 & 9,588,871 & 11,082,963 & 9,588,871 &  - & 8,962,167 & 11,082,963 & 8,962,167 &  - & 125,363,640\tabularnewline
\hline 
\end{tabular}%
}
    \caption{Number of $\sigma$-vectors (top) and number of determinants (bottom) used in solving the linear-response equations in a GS+V+X LR-SCI the $^2J_{\mathrm{HH}}$ spin-spin coupling constant calculation on water/6-311++G{*}{*}-J ($\varepsilon=10^{-4.75}$).}
    \label{tab:si_sigma_vectors}
\end{table}

\newpage
Table \ref{tab:si_timings_vs_cc} reports total execution times for our LR-SCI implementation and for CCSD and CCSDT calculations for the calculation of NMR spin-spin coupling constants of the water molecule.
In this example, it is clear that coupled-cluster methods are more computationally efficient.
We note the difference in core counts used in the two sets of calculations, which translates to (even) larger costs for the LR-SCI when expressed in core-hours.   

\begin{table}
    \centering
\begin{tabular}{|c|c|cc|cc|}
\hline 
 & Basis functions & CCSD (16 cores) & CCSDT (16 cores) & LR-SCI (128 cores) & $\varepsilon$\tabularnewline
\hline 
6-31G-J & 20 & 6 & 807 & 628 & $10^{-4.00}$\tabularnewline
6-31+G{*}-J & 29 & 12 & 3,252 & 10,593 & $10^{-4.75}$\tabularnewline
6-31++G{*}{*}-J & 37 & 20 & 8,552 & 43,554 & $10^{-4.75}$\tabularnewline
6-311++G{*}{*}-J & 46 & 41 & 18,361 & 92,115 & $10^{-4.75}$\tabularnewline
\hline 
\end{tabular}
    \caption{Total wall time (in seconds) required for the calculation of spin-spin coupling constants of water. For the GS+V+X LR-SCI calculations, $\varepsilon$ is adjusted to approximately match a 0.01--0.02 Hz accuracy (similar to CCSDT), except for the calculation with 6-311++G**-J, where a lower 0.05 Hz accuracy was used.}
    \label{tab:si_timings_vs_cc}
\end{table}

\bibliography{main}